\documentclass[]{spie}  
\usepackage[]{graphicx,latexsym}

\newcommand{\etal}{et al.}

\newcommand{\eg}{e.g.}

\newcommand{\micron}{\mbox{$\mu{\rm m}$}}

\newcommand{\Mjup}{\hbox{M$_{\rm Jup}$}}

\newcommand{\arcsec}{\mbox{$^{\prime \prime}$}}

\newcommand{\degs}{\mbox{$^{\circ}$}}
\newcommand{\HST}{{\sl HST}}
\def\lesssim{\mathrel{\hbox{\rlap{\hbox{%
 \lower4pt\hbox{$\sim$}}}\hbox{$<$}}}}
\def\gtrsim{\mathrel{\hbox{\rlap{\hbox{%
 \lower4pt\hbox{$\sim$}}}\hbox{$>$}}}}

\newcommand\aj{AJ}%
\newcommand\apj{ApJ}%
\newcommand\apjl{ApJ}%
\newcommand\aap{A\&A}%
\newcommand\mnras{MNRAS}%
\newcommand\pasp{PASP}%
\newcommand\nat{Nature}%

\newcommand{\twomassbin}{\hbox{2MASS~J1534$-$2952AB}}

\title{LGS AO Science Impact: Present and Future Perspectives}

\author{Michael C. Liu\supit{a}
\skiplinehalf
\supit{a}Institute for Astronomy, University of Hawai`i, 2680
  Woodlawn Drive, Honolulu, HI 96822; USA}

\authorinfo{Email: {\tt mliu@ifa.hawaii.edu}}

\begin{document} 
  \maketitle 

\begin{abstract}
  The recent advent of laser guide star adaptive optics (LGS AO) on
  the largest ground-based telescopes has enabled a wide range of high
  angular resolution science, previously infeasible from ground-based
  and/or space-based observatories.  As a result, scientific
  productivity with LGS has seen enormous growth in the last few
  years, with a factor of $\approx$10 leap in publication rate
  compared to the first decade of operation.  Of the 54~refereed
  science papers to date from LGS AO, half have been published in the
  last $\approx$2 years, and these LGS results have already made a
  significant impact in a number of areas.  At the same time, science
  with LGS AO can be considered in its infancy, as astronomers and
  instrumentalists are only beginning to understand its efficacy for
  measurements such as photometry, astrometry, companion detection,
  and quantitative morphology.  We examine the science impact of LGS
  AO in the last few years of operations, largely due to the new
  system on the Keck~II 10-meter telescope.  We review currently
  achieved data quality, including results from our own ongoing brown
  dwarf survey with Keck LGS.  We assess current and near-future
  performance with a critical eye to LGS AO's capabilities and
  deficiencies.  From both qualitative and quantitative
  considerations, it is clear that the era of regular and important
  science from LGS AO has arrived.
\end{abstract}


\section{INTRODUCTION}
\label{sec:intro}  

Astronomers have envisioned using laser guide star adaptive optics
(LGS AO) to achieve diffraction-limited observations from ground-based
telescopes for over two decades\cite{1985A&A...152L..29F,
  1987Natur.328..229T, 1994OSAJ...11..263H}.  (See
Refs.~\citenum{2004aoa..book.....R} and~\citenum{1998aoat.conf.....H}
for a historical review.)  The realization of this vision has required
the dedication, talents, and resources of myriad individuals and
organizations.
LGS AO is now being used regularly to conduct a broad range of
science.  The promise of near diffraction-limited imaging and
spectroscopy from the ground over most of the sky is coming to
fruition.  This heralds an important new capability for observational
astronomy.

The purpose of this paper is to examine the science that has been done
using LGS AO, with a critical eye to what has been achieved and what
promises and challenges remain.  Much has been accomplished since the
last review of science from LGS (Ref.~\citenum{2006SPIE.6272E..14L}),
which focused on publication output.  Here, we go beyond mere
productivity and examine the science {\em impact} of LGS as
demonstrated in the record of published research.

\section{LGS AO SCIENCE PRODUCTIVITY AND IMPACT}

As of June 2008, five telescopes have generated astronomical science
publications from LGS AO: the 1.5-meter Starfire Optical Range (SOR)
Telescope in New Mexico\cite{1994OSAJ...11..310F}; the 3-meter Shane
Telescope at Lick Observatory in California\cite{max97}; the 3.5-meter
German-Spanish Astronomical Centre Telescope in Calar Alto,
Spain\cite{2000ExA....10....1E}; the 10-meter Keck~II Telescope in
Hawaii\cite{2006PASP..118..297W}; and the 8.1-meter Gemini-North
Telescope in Hawaii \cite{2006SPIE.6272E.114B}.  The Starfire system
used a Rayleigh-backscattered LGS, and the other systems use sodium
laser guide stars.

In the first decade when LGS was available for astronomical science
(1995--2004), science productivity was modest, amounting to 10
refereed papers from Starfire, Lick, and Calar Alto.  In comparison,
as of July 2002, after about a decade of science operation, AO systems
had produced 144 refereed science papers, nearly all derived from
natural guide star (NGS) AO systems.\cite{2003SPIE.4834...84C}.  A
non-comprehensive search of the NASA Astrophysical Data System (ADS)
abstract database in 2006 indicates that the total number of AO
science papers had about doubled since then, again dominated by NGS
observations.  The modest science productivity of LGS AO compared to
NGS AO is no doubt due to the vastly greater complexity and cost of
implementing LGS AO systems.

\begin{figure}[t]
  \begin{center}
    \begin{tabular}{c}
      \includegraphics[width=3.9in,angle=90]{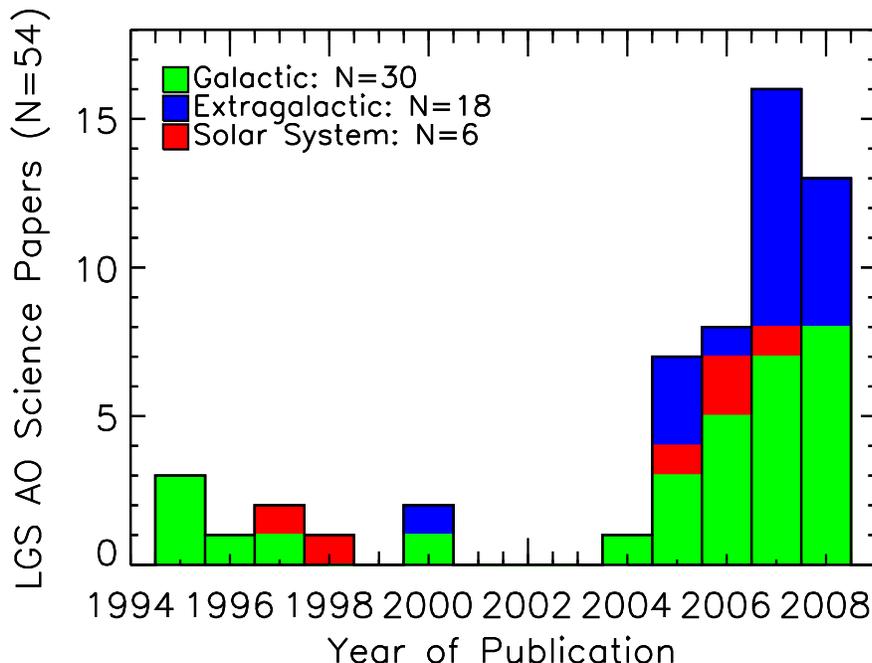}
    \end{tabular}
  \end{center}
  \vskip -0.4in
  \caption[fig:pubs] 
  { \label{fig:pubs} A summary of all refereed LGS AO science papers
    published as of June 2008, 54 in total.  The colors represent
    different science areas.  The large spike in publications since
    2005 comes largely from the LGS AO system on the Keck~II
    Telescope.}
\end{figure} 

However, the last few years have seen enormous growth in LGS science
productivity, as compiled using NASA ADS and summarized in
Figure~\ref{fig:pubs}.  (This compilation includes only papers focused
on astronomical science.  Refereed AO instrumentation, ``first light''
papers, and on-sky atmospheric calibration papers are not included ---
either way, these would amount to a minor perturbation.)  Since 2005,
there has been a surge in refereed publications, largely due to the
sodium LGS AO system on the Keck~II Telescope which accounts for 39
out of the 44 publications from 2005--2008.  Some notable science
highlights from LGS AO include:
\begin{itemize}
\item Identification of small moons around Jupiter Trojan asteroids
  and Kuiper Belt Objects and subsequent direct density determinations
  by Marchis \etal\ (Ref.\ \citenum{2006Natur.439..565M}) and Brown
  \etal\ (Refs.\ \citenum{2005ApJ...632L..45B, 2006ApJ...639L..43B});
\item Measurement of the atmospheric properties and dynamical masses
  of brown dwarfs by Liu \etal\ (Refs.\ \citenum{2005astro.ph..8082L,
    2006astro.ph..5037L, liu08-2m1534orbit} and \S~3);
\item Discovery of spatially resolved photoevaporating gas structures
  around young pre-main sequence stars in the Orion Nebula Cluster by
  McCullough \etal\ (Ref.\ \citenum{1995ApJ...438..394M});
\item Examination of the extended and/or time-variable emission around
  the Galactic Center by Ghez \etal\ (Ref.\
  \citenum{2005ApJ...635.1087G, 2007ApJ...667..900H}) and Krabbe
  \etal\ (Ref.\ \citenum{2006ApJ...642L.145K});
\item Determination of the mass function and the proper motion of the
  Arches Cluster in the vicinity of the Galactic Center by Kim \etal\
  (Ref.\ \citenum{2006ApJ...653L.113K}) and Stolte \etal\ (Ref.\
  \citenum{2008ApJ...675.1278S});
\item Identification of the progenitor stars of Type~II supernovae by
  combining Keck LGS imaging of the SN with pre-explosion images in
  the \HST\ Archive by Gal-Yam \etal\ (Refs.\
  \citenum{2005ApJ...630L..29G, 2007ApJ...656..372G});
\item Measurement of the stellar velocity dispersion of a luminous
  quasar host galaxy by Watson \etal\ (Ref.\
  \citenum{2008arXiv0806.3271W});
\item Measurement of the spatially resolved kinematics of $z=1.5-3$
  star-forming galaxies by Wright, Law \etal\ (Refs.\
  \citenum{2007ApJ...658...78W,2007ApJ...669..929L}).
\end{itemize}
\noindent 
Except for the Orion study which was done with the 1.5-meter Starfire
Telescope and the quasar host galaxy result from the Gemini-North
8.1-meter Telescope, all of these results come from the Keck LGS
system.  This is not surprising, given this is the first facility LGS
AO system on an 8--10~meter class telescope and has been in regular
operation for 3+ years.  As such, it provides a good illustration of
the areas of science which astronomers are eager to explore using LGS.
To illustrate the wide diversity of topics, {\bf Table~1} summarizes
the science areas that have been addressed with the Keck system so
far.  Galactic science provides the majority of the science, and
within that category, half of the papers relate to brown dwarfs and/or
low-mass stars.  However, a significant fraction ($\approx$40\%) of
papers are in the area of extragalactic science, a notable (and
anticipated) change from science done with NGS.

\begin{table}[h]
  \caption{{\bf Keck LGS AO science papers, sorted by science
      topic (39 total papers).}} 
\begin{center}       
\begin{tabular}{c|c|l}
\rule[0ex]{0pt}{3.5ex}  
Area & Number of Papers & Sub-Topic (Number of Papers) \\
\hline
\hline

\rule[-0.5ex]{0pt}{3.5ex}  
Galactic & 20 & Brown dwarfs and low-mass stars (N=11) \\
         &    & Galactic Center (N=6) \\
         &    & Compact objects (N=2) \\
         &    & Star formation (N=1) \\  \hline

\rule[-0.5ex]{0pt}{3.5ex}  
Extragalactic & 15  & High-redshift galaxies (N=6) \\
              &     & Gravitational lensing (N=3) \\
              &     & Stellar populations (N=3) \\
              &     & Supernovae (N=3) \\ \hline

\rule[-0.5ex]{0pt}{3.5ex}  
Solar System & 4   & Kuiper Belt (N=3) \\
             &     & Asteroids (N=1) \\ 

\hline
\end{tabular}
\end{center}
\end{table}

While the quantity of publications is an unambiguous metric of science
output, a more difficult issue to {\em quantify} is whether the LGS AO
science being done has been scientifically important/impactful.  As
one (and only one!) line of analysis, we examine the citation counts
of LGS science papers in light of the study by
Pearce\cite{2004A&G....45b..15P}.  He has assembled the distribution
of citation counts as a function of year since publication, allowing
us to derive a citation ranking of LGS science papers relative to all
astronomical science papers published at the same time.  {\bf Table~2}
presents the top cited LGS science papers, both in terms of total
citations and relative ranking.  Overall, a number of recent papers
have been in the top 1--3\% of citations relative to their
contemporary publications.  (In the spirit of measuring LGS science
impact, the Table only included papers whose science content was
substantially based on LGS observations, not ones where the LGS data
play an inconsequential role.  This criterion excludes one highly
cited paper, Ref.~\citenum{2007ApJ...666.1116S}, which would have
placed 2nd on the list with 34 citations in one year since
publication.)

\begin{table}[t]
  \caption{{\bf Top-cited science papers from LGS AO.}  Some paper titles have been abbreviated. 
    ``N(cite)'' gives the number of citations, based on data from NASA
    ADS.  ``Rank'' gives the citation counts of each paper relative to
    other astronomy papers published at the same time based on the
    compilation of Pearce (Ref.~\citenum{2004A&G....45b..15P}), \eg, 99\% means
    that the paper is more highly cited than nearly all papers published
    at the same time. ``Field'' gives the area of study: ``SS'' = solar
    system, ``Gal'' = galactic, ``Xgal'' = extragalactic.  ``$N_{obj}$''
    indicates the number of science targets/fields observed with LGS.
    Note that NASA ADS is incomplete for citations
    of solar system papers, hence the ``$>$'' for the last paper.} 
\begin{center}       
\vskip -3ex
\begin{tabular}{c|p{0.8cm}|c|p{3.2in}|c|c|c|c} 
\rule[0ex]{0pt}{3.5ex}  
N(cite) & Rank & Year & Authors: Title, Journal & Tel & Field &
$\lambda\lambda$ & $N_{obj}$ \\
\hline
\hline

\rule[-0.5ex]{0pt}{3.5ex}  
42  & 90\% & 1995  & {\bf \em McCullough \etal,}
      {\em Photoevaporating Stellar Envelopes Observed with Rayleigh
        Beacon AO}, ApJ
     & SOR & Gal & H$\alpha$ & 1 \\ \hline  

\rule[-0.5ex]{0pt}{3.5ex}  
30  & 99\% & 2007  & {\bf \em Gal-Yam \etal,}
      {\em On the Progenitor of SN 2005gl and the Nature of Type IIn
        SN}, ApJ
     & Keck & Xgal & $K^\prime$ & 1 \\ \hline  

\rule[-0.5ex]{0pt}{3.5ex}  
30  & 98\% & 2006  & {\bf \em Liu \etal,}
      {\em A Novel T~Dwarf Binary and the Potential Role of Binarity
        in the L/T Transition}, ApJ
     & Keck & Gal & $JHK$ & 1 \\ \hline  

\rule[-0.5ex]{0pt}{3.5ex}  
30  & 97\% & 2005  & {\bf \em Ghez \etal,}
      {\em LGS AO of the Galactic Center: Sgr A*'s IR Color and its
        Extended Red Emission}, ApJ
     & Keck & Gal & $KL^\prime$ & 1 \\ \hline

\rule[-0.5ex]{0pt}{3.5ex}  
30  & 97\% & 2005  & {\bf \em Liu \etal,}
      {\em Kelu-1 is a Binary L~Dwarf: First Brown Dwarf Science with
        LGS AO}, ApJ
     & Keck & Gal & $JHK$ & 1 \\ \hline

\rule[-0.5ex]{0pt}{3.5ex}  
$>$18 & $>$97\% & 2006  & {\bf \em Brown \etal,}
        {\em Satellites of the Largest Kuiper Belt Objects}, ApJ
      & Keck & SS & $K^\prime$ & 4 \\

\hline
\end{tabular}
\end{center}
\end{table}

\section{SUMMARY OF KECK LGS PERFORMANCE}

\subsection{A High Angular Resolution Survey of Field Brown Dwarfs}

To illustrate what is currently possibly with LGS, we examine the
on-sky performance of the Keck LGS system.  Since the inception of the
Keck system for open science use in 2005, my collaborators and I have
been conducting a high angular resolution near-IR study of nearby
brown dwarfs.  Our goals are (1) to assess the binary frequency of
ultracool dwarfs; (2) to test atmospheric models with these coeval
systems, \eg, as associated with the abrupt spectral transition from
the L~dwarfs to the T~dwarfs;\cite{2006astro.ph..5037L} (3) to search
for exceptionally low-temperature
companions;\cite{2008A&A...482..961D} and (4) to find and monitor
substellar binaries suitable for dynamical mass
determinations.\cite{liu08-2m1534orbit}

LGS AO represents a major advance for this science area.  Most known
brown dwarf binaries have separations of
$\lesssim$0.3\arcsec,\cite{2006astro.ph..2122B} hence the need for
high angular resolution imaging to find and characterize them via
resolved photometry and spectroscopy, but they are too optically faint
for natural guide star AO.  Our observations have achieved
3--4$\times$ the angular resolution at $K$-band (2.2~\micron) compared
to \HST\ and thus are more sensitive to close companions.  In
addition, the ability of Keck LGS AO to find tighter binaries means
that systems with much shorter orbital periods than the current sample
can be found and expeditiously monitored.  Finally, many of the key
spectral diagnostics of brown dwarfs are in the infrared and thus
probing the atmospheric properties of these objects with resolved
colors and spectra of binaries is well-suited to current LGS
capabilities.

\begin{figure}[h]
  \vskip -0.5in
  \begin{center}
    \begin{tabular}{c}
      \hskip -0.7in
      \includegraphics[width=3.5in,angle=90]{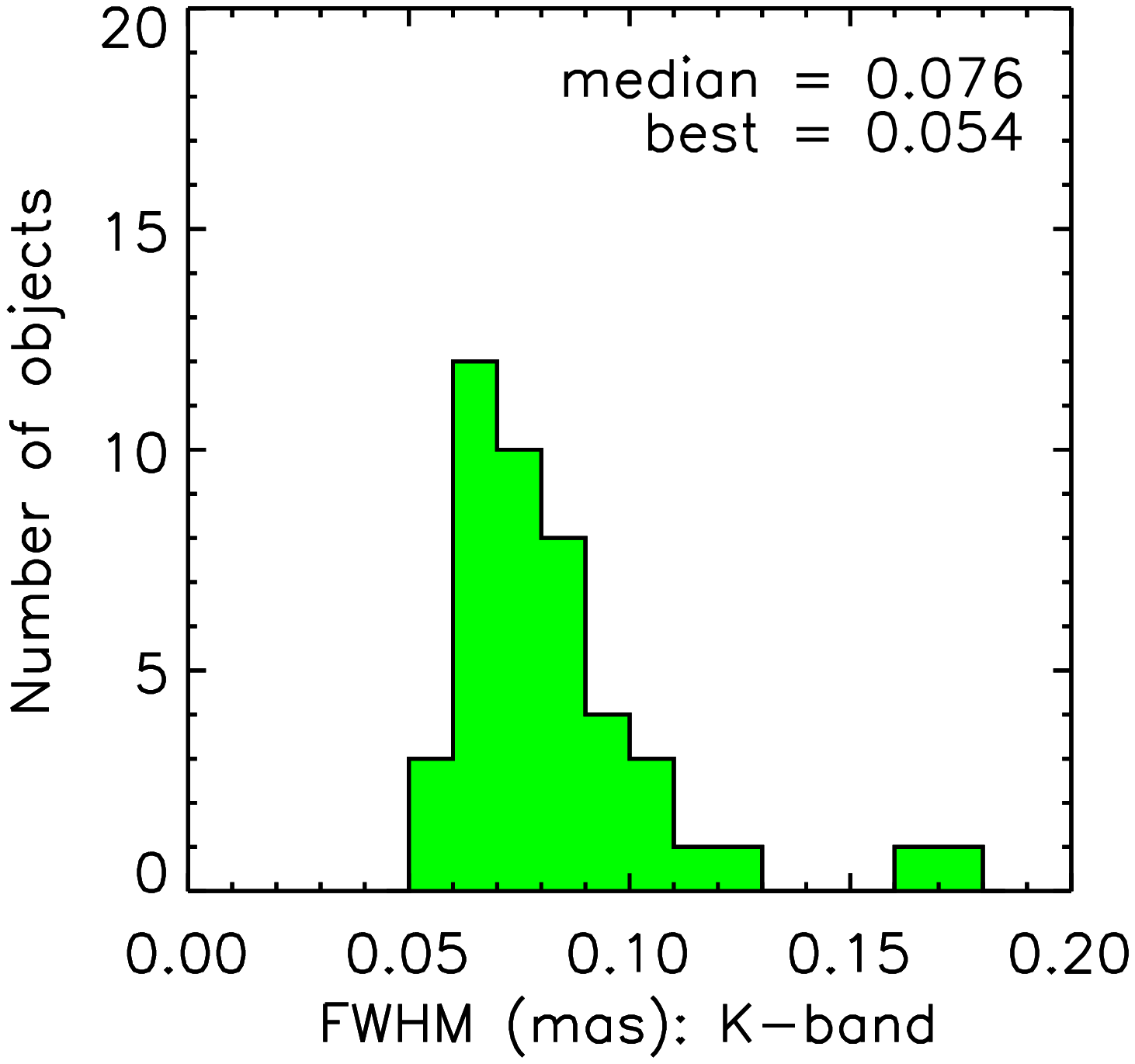}
      \hskip -1.6in
      \includegraphics[width=3.5in,angle=90]{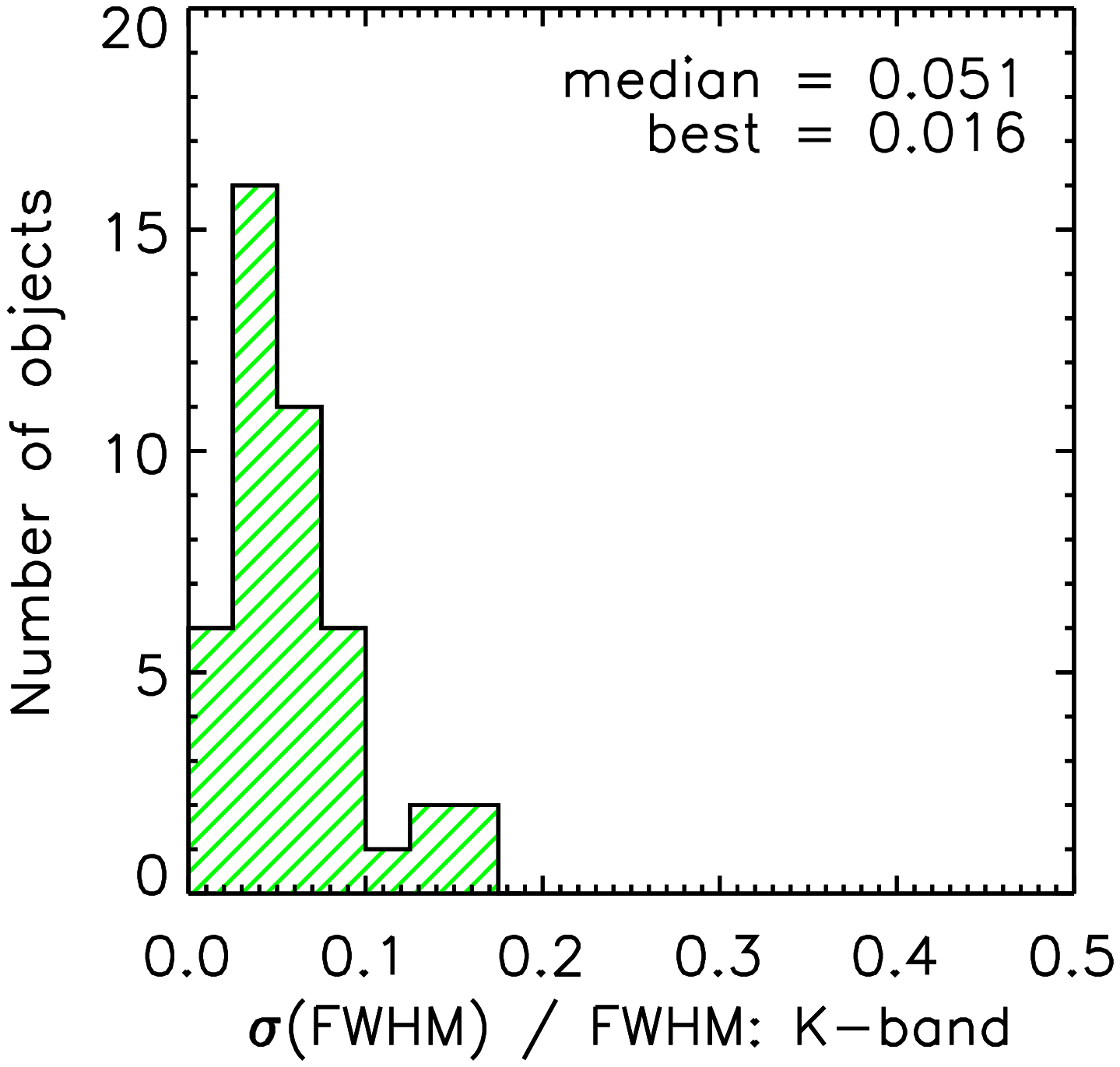}\\
      \hskip -0.6in
      \includegraphics[width=3.5in,angle=90]{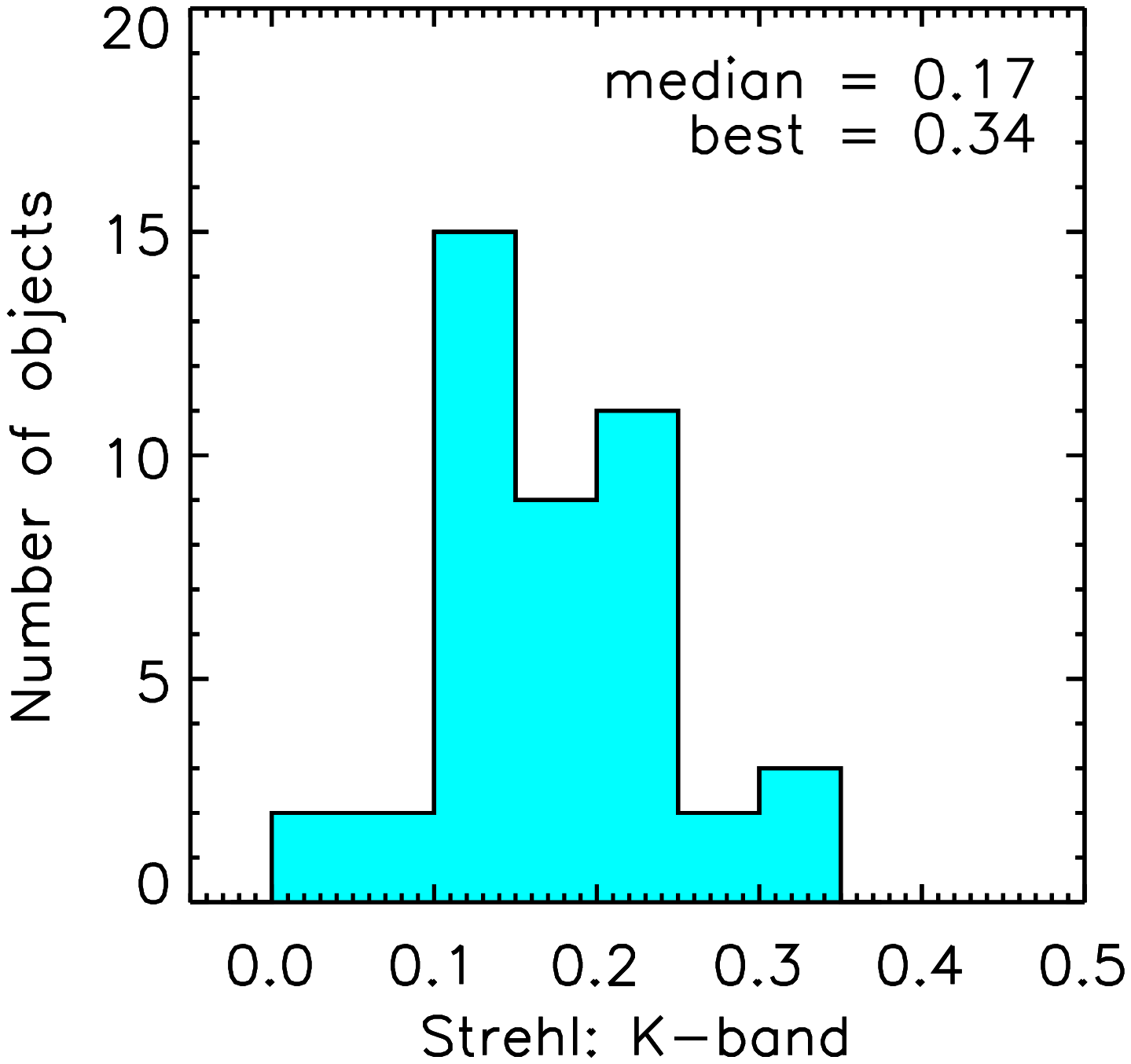}
      \hskip -1.6in
      \includegraphics[width=3.5in,angle=90]{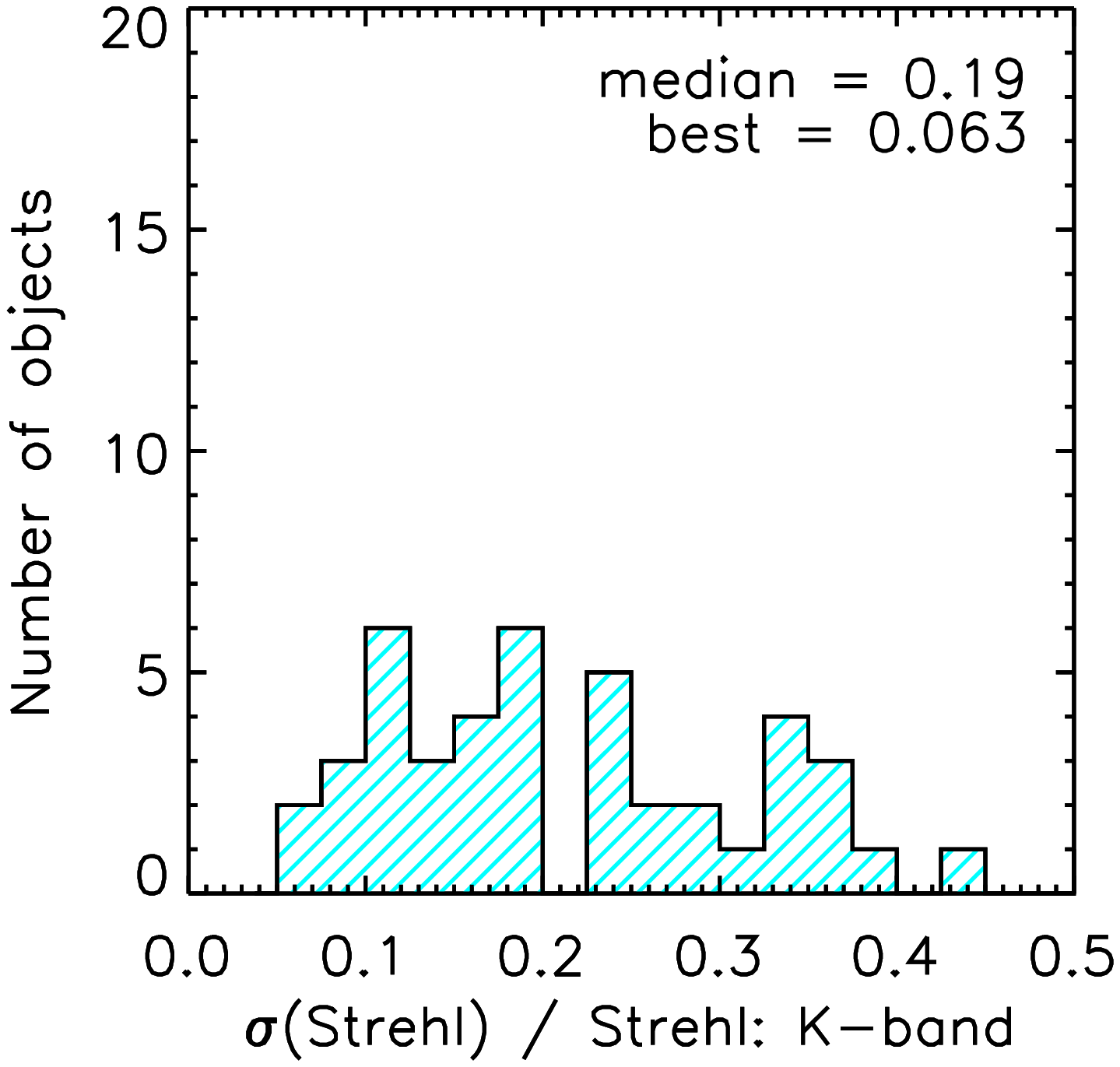}\\
    \end{tabular}
  \end{center}
  \vskip -4ex
  \caption[fig:kecklgs] 
  { \label{fig:kecklgs} Summary of Keck LGS AO $K$-band (2.2~\micron)
    performance, based on our near-IR imaging survey of brown dwarfs
    (Liu \etal, in prep).  No bad data have been censored, so the
    compilation comprises a mix of seeing conditions, target
    airmasses, and technical performance.  {\bf Top panels:} Histogram
    of $K$-band FWHM and its fractional RMS variation within a given
    dataset.  Typically each dataset is composed of 6--12 individual
    images taken over an elapsed time of 10--20~minutes.  The median
    FWHM is 0.076\arcsec\ with a median RMS variation of 5\%.  {\bf
      Bottom panels:} Same histograms for $K$-band Strehl.  The median
    Strehl is 0.17 with an RMS variation of a factor of 1.19 (\eg,
    Strehl = $0.17\pm0.03$).}
\end{figure}

Our brown dwarf imaging survey provides an excellent dataset for
assessing typical Keck LGS performance in the case of off-axis
observations, namely the situation where the LGS is pointed to the
science target but tiptilt sensing and correction are derived from an
adjacent field star.  Brown dwarfs are far too optically faint to
serve as their own tiptilt references and hence the need for a nearby
tiptilt star -- this is the same observing situation as for many
extragalactic LGS applications and thus provides a good reference
point.  For Keck, the tiptilt star must be within 60\arcsec\ of the
science target -- in practice, we find that this results in a sky
coverage fraction of about 2/3 for an estimated $K$-band Strehl ratio
of $\gtrsim$0.2.  Since targets for our brown dwarf program span most
of the sky (except for avoidance of the galactic plane), this 2/3 sky
coverage estimate is a fair representation of the fraction of any set
of generic targets that can be imaged with LGS.

Figure~\ref{fig:kecklgs} summarizes the image quality of a subset of
our Keck LGS observations of nearby brown dwarfs, spanning multiple
observing runs from 2005--2007.  No bad data have been censored, so a
mix of seeing conditions, off-axis tiptilt star properties, and
technical performance (e.g., LGS projected power and sodium light
return flux) are represented.  (See also Ref.\
\citenum{2006SPIE.6272E..14L} for more performance descriptions.)  The
median $K$-band image FWHM for our survey is 0.076\arcsec\ with a best
value of 0.051\arcsec.  The median $K$-band Strehl is 0.17 with a best
value of 0.34.

LGS images are naturally time-variable, and the shape and detailed
structure of the PSF changes in every image.  Figure~\ref{fig:kecklgs}
provides one representation of this PSF variability, plotting
histograms of the fractional RMS deviations in $K$-band FWHM and
Strehl ratios.  Typically, each of our brown dwarf datasets
constitutes a series of 6--12~images, each with a integration of about
1~min and total elapsed time of about 15--30~min.  Over these time
scales, the plotted histograms show significant FWHM and Strehl
variations.  This is obviously a challenge for science programs
requiring a stable PSF.

\begin{figure}[t]
  \vskip -0.3in
  \begin{center}
    \begin{tabular}{c}
      \hskip -0.75in
      \includegraphics[width=3.75in,angle=90]{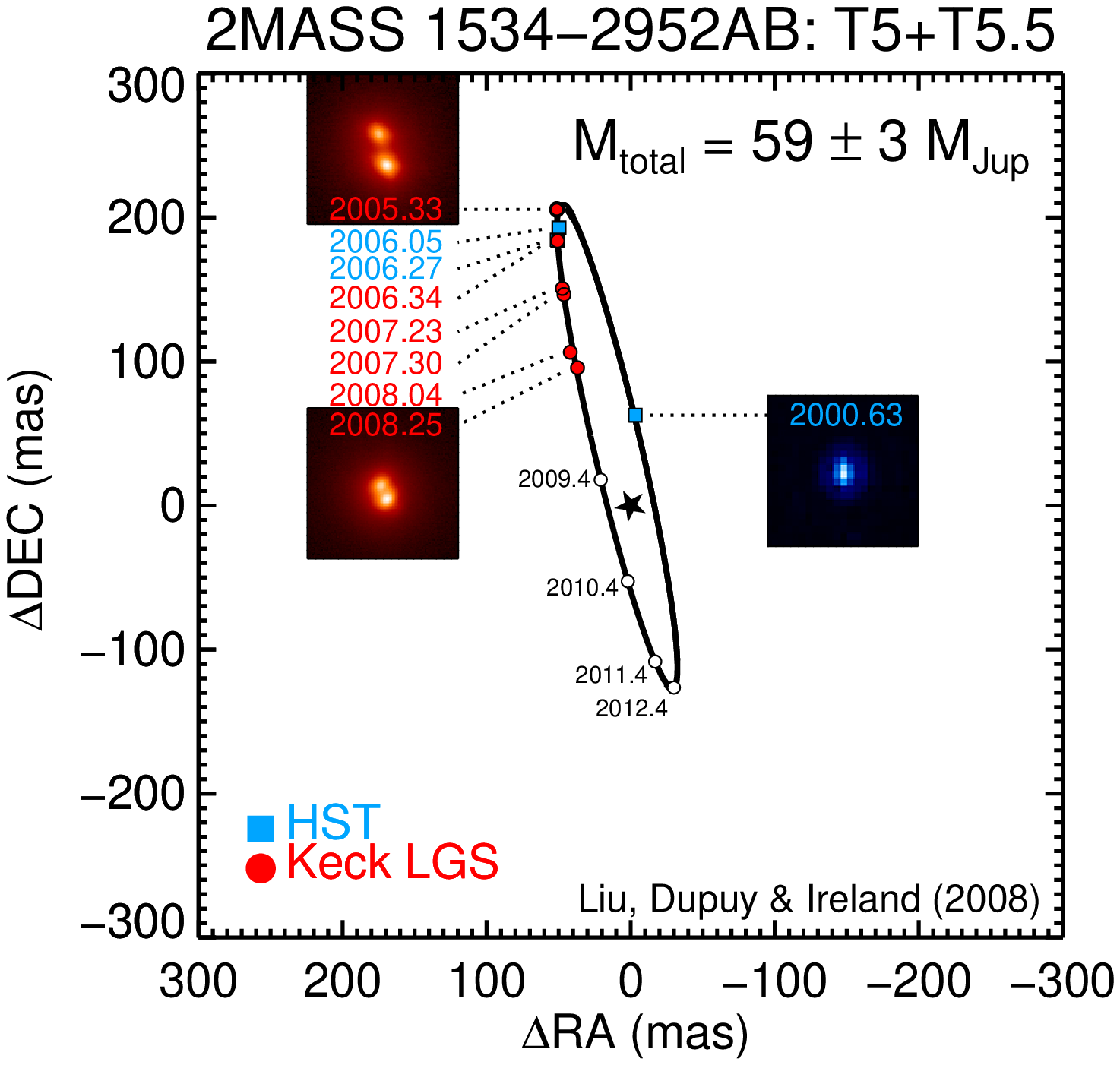}
      \hskip -1.8in
      \includegraphics[width=3.75in,angle=90]{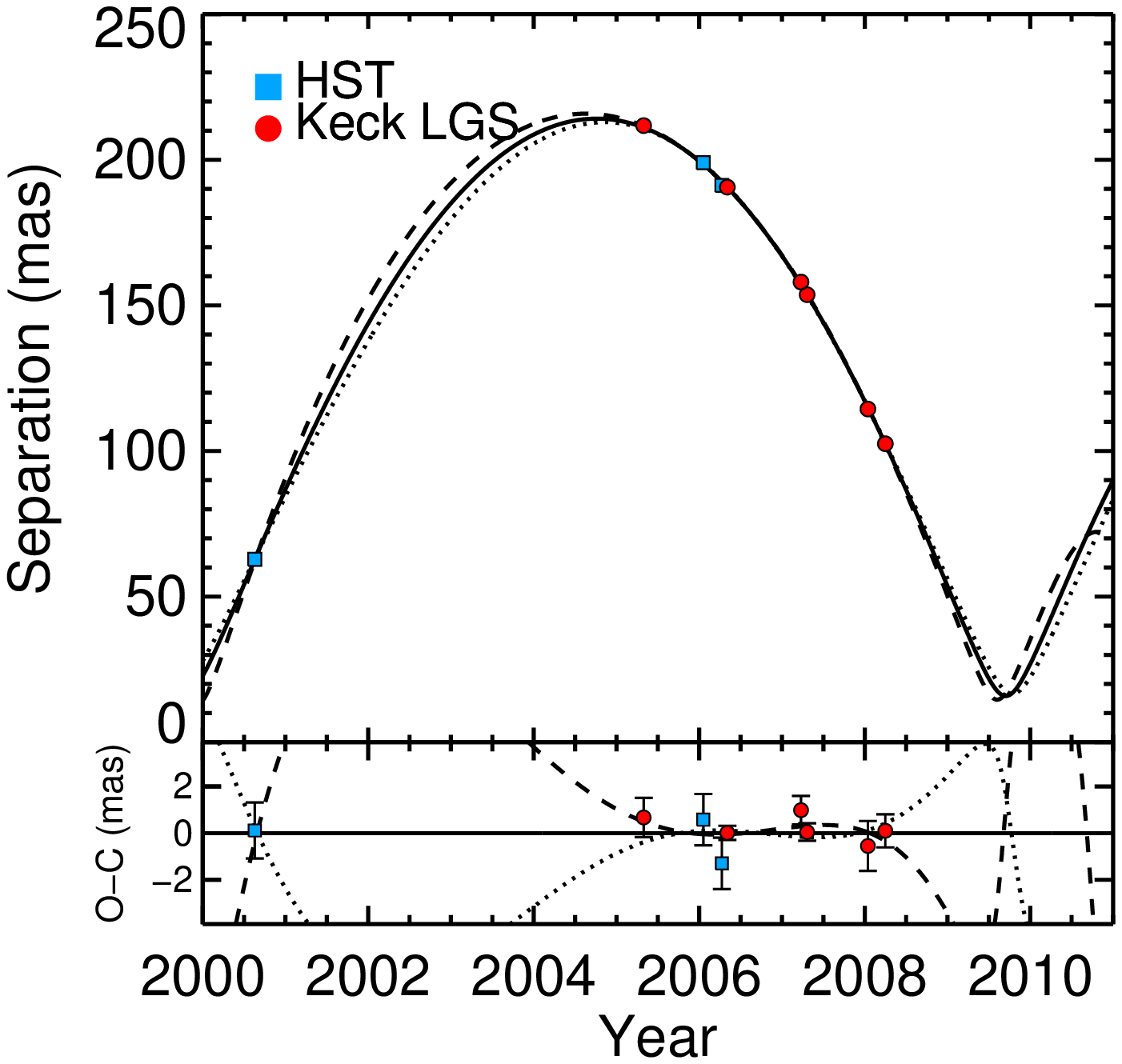}
    \end{tabular}
  \end{center}
  \vskip -6ex
  \caption[fig:2m1534-orbit] 
  { \label{fig:2m1534-orbit} First dynamical mass determination for a
    binary T~dwarf, the T5.0+T5.5 binary 2MASS~J1534-2952AB (Ref.\
    \citenum{liu08-2m1534orbit}).  Six epochs of data from Keck LGS
    are combined with \HST\ archival data to form a dataset from
    2000-2008 that spans about 50\% of the 15-year orbital period.
    Through careful analysis of the binary images with PSF fitting,
    the Keck LGS data achieve sub-milliarcsecond relative astrometry
    of the two components.  The total mass is $59\pm3~\Mjup$.  This is
    the first field binary for which both components are directly
    confirmed to be below the stellar/substellar limit
    ($\lesssim$75~\Mjup).  This is also the coolest and lowest mass
    binary with a dynamical mass determination to date. The left plot
    shows the relative astrometry and best-fitting orbit.  The right
    plot shows the separation measurements compared to the best
    fitting orbit (solid line) and two alternative orbital periods of
    similar total mass (dashed and dotted lines).  The right bottom
    sub-panel shows the residuals between the observations and the
    predictions from the orbits.}
\end{figure}

Despite the notable PSF fluctuations, high precision science can be
with Keck LGS data.  For example, we have been using LGS to carefully
monitor the orbits of binary brown dwarfs in order to determine their
dynamical masses and thereby test theoretical models.  Despite the
hundreds of brown dwarfs that have been identified and had their
spectrophotometric properties characterized, direct measurements of
their most fundamental property, namely their mass, are sorely
lacking.  Typical estimated orbit periods are a decade a more, but
many of the binaries were discovered $\gtrsim$5~years ago with \HST\
and thus multi-epoch LGS followup is well-suited to dynamical mass
determinations.

Figure~\ref{fig:2m1534-orbit} shows an example of what is possible,
presenting the orbit of the T5+T5.5 binary 2MASS~J1534-2952AB by Liu,
Dupuy and Ireland (Ref.\ \citenum{liu08-2m1534orbit}). This is the
first dynamical mass determination for a binary T~dwarf, the coolest
and least luminous class of brown dwarfs.  There are two stars within
6\arcsec\ of \twomassbin\ which we use as PSFs to deblend the light of
the two binary components.  Through extensive Monte Carlo testing of
different deblending methods applied to simulated images of binary
stars constructed from the single PSFs, we are able to measure the
relative position and orientation of the components to 0.3--0.7~mas
and 0.2--0.7\degs\ RMS at $K$-band.  The precision of the best
measurements is such that atypical sources of uncertainty need to be
considered, including:
\begin{enumerate}
\item The calibration of the instrumental plate scale and orientation
  can introduce additional uncertainty, since these values for the
  Keck facility near-IR camera are known to about 1~part in $10^{-3}$
  and 0.1\degs, respectively.
\item The differing spectral types of the two components means that
  the light from each is subjected to a slightly different amount of
  differential chromatic refraction (DCR).  Given the sky orientation
  of the binary (just about North-South), the declination of the
  target ($\delta=-29\degs$ meaning the smallest possible airmass for
  Keck observing is 1.55), and the fact that we observe it near
  transit (for best AO performance), the DCR causes the separation of
  the binary to appear slightly smaller at $H$ and $K$-bands and
  slightly larger at $J$-band compared to the true position as would
  be observed at zenith.  The amplitude of the DCR effect is about
  0.3~mas, much smaller than the measurement errors at most (but not
  all) of the Keck epochs.  However, the effect is a systematic one so
  we correct the relative astrometry of the two components based on
  synthetic photometry from T~dwarf spectra.  (Since T~dwarfs have
  suppressed flux at the longward portion of the $K$-band, the DCR is
  smaller than it would be for L~dwarfs observed at the same airmass.)
\end{enumerate}
\noindent The quality of the Keck LGS astrometry equals or exceeds
that available for the same target from \HST, though at the very
smallest separations ($\approx$FWHM) \HST\ data still have an
advantage since the very stable \HST\ PSF can be simulated to high
fidelity\cite{1993ASPC...52..536K} and then fitted to tight binary
images (\eg, Ref.\ \citenum{liu08-2m1534orbit}).  However, LGS AO
provides the necessary long-term platform for synoptic monitoring of
visual binaries, especially since the required amount of observing
time at each epoch is relatively modest but many epochs are needed.
This in contrast to \HST\ where target acquisition is slow and
monitoring a populous sample over many epochs is telescope
time-intensive.


\subsection{Review of Published Keck Performance}

To further summarize the current capability, {\bf Table~3} provides a
summary of achieved Keck LGS performance as presented in the published
science papers.  Obviously this is a heterogeneous assemblage from both
the standpoint of the observations and the subsequent analysis, but it
does provide ``ground truth'' of the quality of on-sky data that has
proven suitable for publication.  We consider several types of
measurements that readily lend themselves to quantitative
uncertainties (as opposed to more difficult measurements to quantify
such as, \eg, morphological studies of resolved objects):
\begin{itemize}
\item {\em Relative photometry:} the flux of a science target relative
  to other sources in the same image of known brightness (\eg, from
  2MASS);
\item {\em Absolute photometry:} the absolute flux of a science target as
  directly measured from the images of the target and then compared to
  a photometric calibrator star observed contemporaneously (but not
  simultaneously);
\item {\em Relative astrometry:} the location of sources in an LGS
  dataset relative to each other.
\item {\em Absolute astrometry:} the location of sources in a LGS
  dataset as referenced to absolute astrometry tied to an external
  dataset (\eg, \HST).
\item {\em Binary (relative) photometry:} the relative flux ratio of a
  binary;
\item {\em Binary (relative) astrometry:} the separation and position
  angle of a binary;
\item {\em Crowded field astrometry:} the positions and/or proper
  motions for a field with many sources (\eg, star clusters and
  resolved nearby galaxies);
\item {\em Crowded field (relative) photometry:} the same as relative
  photometry above, except for a field with many sources;
\item {\em Crowded field (absolute) photometry:} the same as absolute
  photometry above, except for a field with many sources.
\end{itemize}

\noindent {\bf Table~3} shows that high quality measurements have been
achieved with LGS, especially for relative measurements.  In the
published papers, the most measurements are available for relative
photometry and astrometry of low-mass binaries, since this is a class
of science that readily benefits from LGS.



\section{CHALLENGES AND SOLUTIONS }

\begin{figure}[t]
  \begin{center}
    \vskip -0.2in
    \begin{tabular}{c}
      \hskip -0.5in
      \includegraphics[width=2.8in,angle=90]{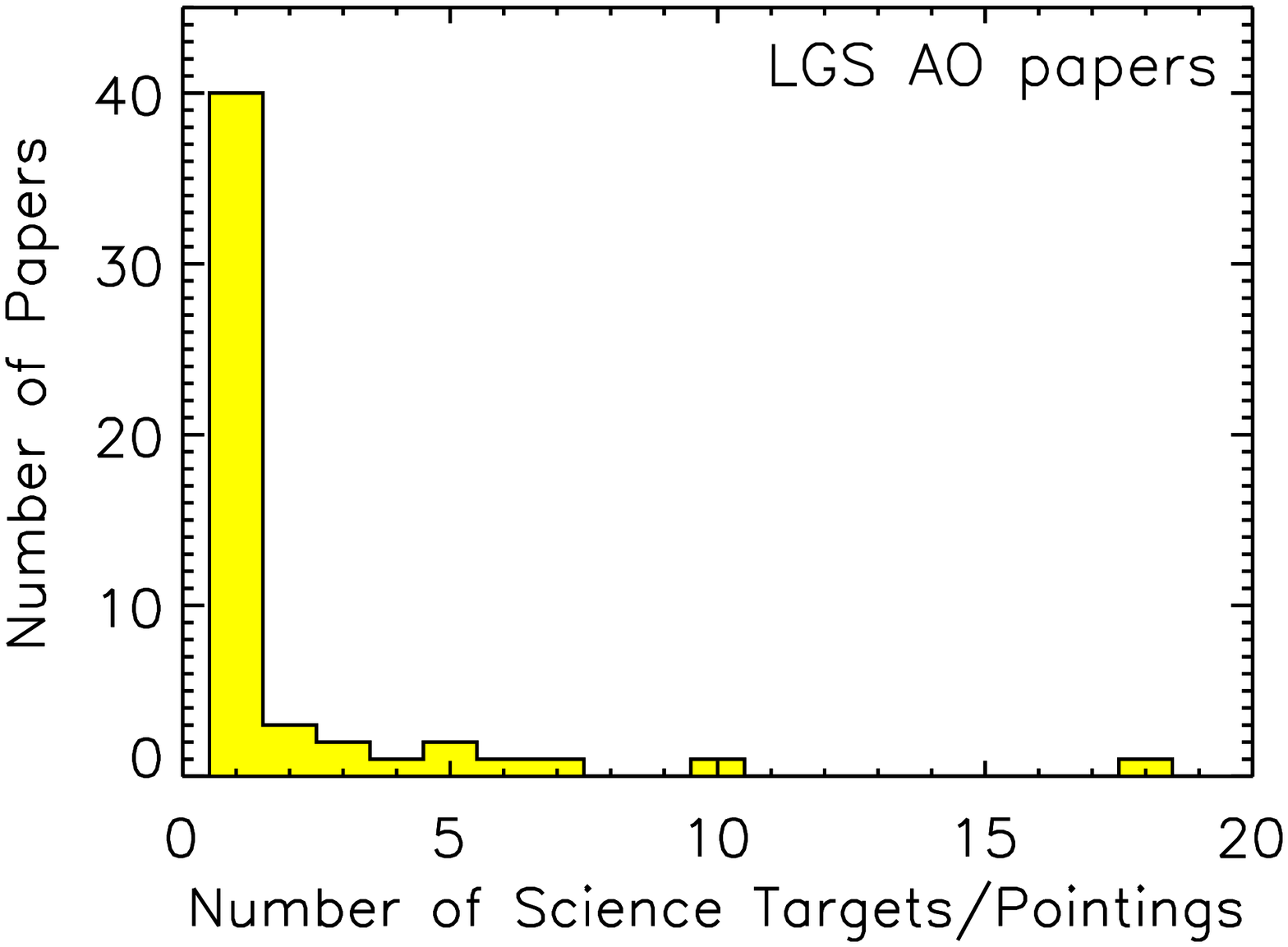}
      \hskip -0.5in
      \includegraphics[width=2.8in,angle=90]{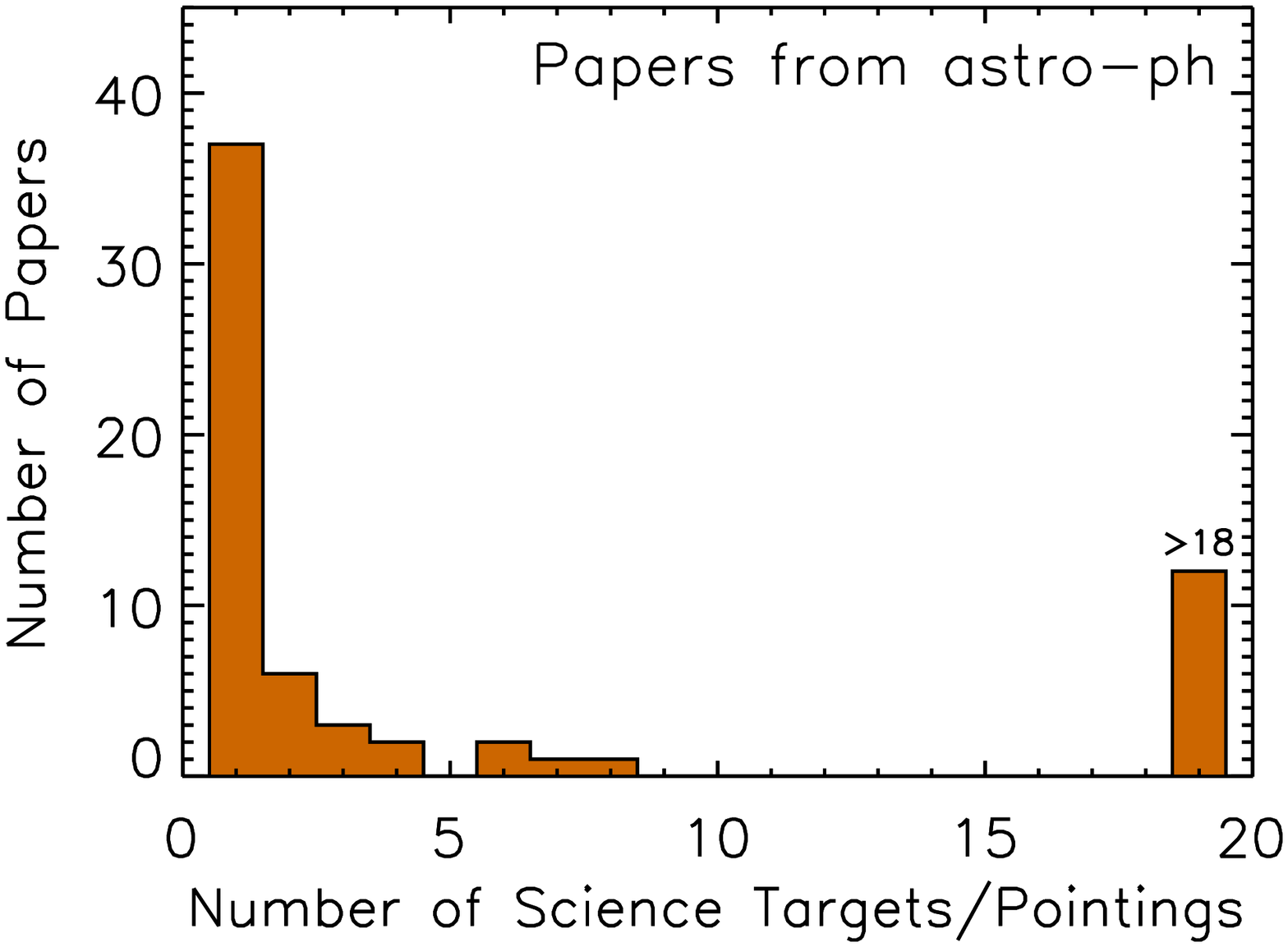}
    \end{tabular}
  \end{center}
  \vskip -0.3in
  \caption[fig:objectcounts] 
  { \label{fig:objectcounts} A histogram of the number of science
    targets and/or pointings observed in LGS AO science papers ({\em
      left}) compared to a comparable number of observational papers
    from randomly chosen astro-ph daily email notifications ({\em
      right}).  In general, the vast majority of papers in both
    samples are focused on a small number of objects, though LGS
    papers to date have not focused on more populous samples of
    objects compared to the astro-ph sample. A two-sided
    Kolmogorov-Smirnov test shows only a 17\% chance that the two
    distributions are drawn from the same sample.}
\end{figure}

There are obviously still a number of challenges to be overcome in
order for LGS AO be considered a regular component of every
astronomer's ``toolkit.''
As another examination of science being done with LGS,
Figure~\ref{fig:objectcounts} shows a histogram of the number of
science targets and/or (substantially different) pointings in an LGS
paper --- most LGS papers ($77\%\pm12$\%) are focused on a single
object/pointing.  For reference, we plot to the same histogram for a
sample of observational papers selected from randomly chosen astro-ph
daily email notifications from 2006.  The distribution of the number
of targets/fields is similar for both the LGS and astro-ph samples, in
that most papers are focused on a small number of objects.  However, a
two-sided Kolmogorov-Smirnov test shows only a 17\% probability that
the two samples are drawn from the same parent population.  The LGS
sample has a somewhat larger fraction of papers based on only a single
object/field.  Also, while the most populous survey in any LGS paper
is composed of 18 objects,\cite{2007ApJ...668..507L} the astro-ph
sample has $19\%\pm5$\% of its papers containing more than 18 objects.
The differences in the two samples is no doubt due to several factors,
including (1) the present technical difficulties of using LGS for
``survey-style'' science, (2) conservative planning on the part of new
LGS observers, and/or (3) the desire to promptly observe notable
objects with a brand-new observational capability (a.k.a.,
``low-hanging fruit'').

As seen in the published work to date, a number of outstanding
challenges exist for carrying science measurements.  An incomplete
list includes:

\begin{itemize}

\item {\bf Relative photometry and astrometry at small separations:}
  Many of the Keck LGS papers are concerned with scenes that involve
  measuring relative fluxes and/or positions of sources at small
  separations, \eg, brown dwarf binaries, the Galactic Center,
  gravitational lenses, and resolved stellar populations.  There are
  substantial gains to be realized from reduction in the measurement
  uncertainties, \eg, by increasing the stability of the LGS AO PSF
  and/or knowledge of the real-time achieved PSF.  For instance, for
  binary measurements LGS has achieved as good as $\approx$0.3~mas,
  but more accurate astrometry would lead to more accurate dynamical
  masses with a shorter time baseline.

  From the standpoint of photometry, most science programs require a
  bare minimum of $\approx$0.1~mag photometric uncertainty, \eg,
  since the IR colors of stars span only a few tenth's of a
  magnitudes.  A level of $\approx$0.03~mag is sufficient for many
  quantitative applications, \eg, spectral type estimates for binaries
  and color-magnitude diagrams of resolved stellar populations, and
  LGS AO has already demonstrated this level of accuracy.  But it has
  yet to achieve the $\lesssim$0.01~mag level routinely achieved with
  modest effort from the ground and space.

\item {\bf Absolute photometry in sparse fields:} None of the
  published Keck LGS papers attempt to do direct photometric
  calibration of imaging data, but rather calibrate all data against
  existing seeing-limited data.

\item {\bf Instrumental astrometric calibration:} About half the Keck
  LGS papers involve high precision astrometry, for programs involving
  solar system bodies, brown dwarf binaries, the Galactic Center, and
  SN progenitors, and these programs are likely to continue to be
  important long-term science.  The current Keck facility camera
  calibration has been adequate, but improving this to a level of
  $\lesssim10^{-3}$ in pixel scale and $\lesssim0.1\degs$ would be
  valuable.

\item {\bf PSF variability and uncertainty:} Significant quantitative
  science has been done {\em despite} the temporal and spatial
  variability of the LGS AO PSF.  In some cases, the science programs
  benefited from having one or more PSF stars in the same field as the
  science target; this is especially valuable for programs needing
  precise measurement such as astrometric monitoring or quantitative
  analysis of complex morphologies (\eg, Refs.\
  \citenum{2005ApJ...635.1087G, 2007ApJ...671.1196M,
    2008ApJ...675.1278S, liu08-2m1534orbit}).  In other cases, PSF
  images taken near-contemporaneously have been used for modeling;
  while less optimal, this may be suitable for programs where
  basically the average bulk properties of the PSF (\eg, encircled
  energy as a function of radius) are more important than the detailed
  structure, such as high-redshift galaxies where the relevant angular
  scales are larger than diffraction-limited PSFs.  The reliability of
  such techniques can be validated through careful modeling of PSF
  effects on the science data (Ref.~\citenum{2008AJ....135.1207M}).
  One example of this has been the work of Sheehy and collaborators
  (Ref.~\citenum{2006astro.ph..4551S}), who tackle a specific PSF
  aspect needed for their science program, namely the determination of
  aperture corrections for crowded field photometry.  Similarly,
  studies by Steinbring \etal, Britton, Cresci \etal, and Cameron
  \etal\ (Refs. \citenum{2005PASP..117..847S, 2006PASP..118..885B,
    2006A&A...458..385C, 2008arXiv0805.2153C}) have addressed methods
  to account for PSF spatial variations in NGS AO data and seem
  valuable for extending to LGS AO data.

\item {\bf Non-redundant aperture masking:} one promising technique is
  the application of non-redundant aperture masking to LGS
  observations.  By replacing the usual clear pupil of the science
  instrument with a milled plate of holes that form a series of
  non-redundant baselines, atmospheric and AO noise can be largely
  eliminated.  This enables interferometric-style imaging over a small
  FOV but to very tight separation ($\lesssim\lambda/D$).  Aperture
  masking has been amply used with Keck NGS AO (\eg,
  Refs.~\citenum{2000PASP..112..555T, 2001ApJ...562..440D}).  The
  first application of this technique with LGS AO is presented by
  Burgasser, Liu, Ireland \etal\ (Ref.~\citenum{2008arXiv0803.0295B}),
  where aperture masking was used to obtain $\approx$3~mag gain in
  contrast over LGS direct imaging to search for faint companions at
  radii $\lesssim 2 \times$FWHM.  The limitation of the method is that
  $\approx$90\% of the instrument pupil plane is blocked out, meaning
  the method is restricted to IR bright objects.  However, it does
  provide a promising method for achieving much greater contrast at
  small separations compared to direct imaging.

\end{itemize}

Since the previous review on LGS science (Ref.\
\citenum{2006SPIE.6272E..14L}), it is clear that LGS science has
leaped forward.  The last few years have witnessed a veritable boom in
science from LGS, as seen in a doubling of science papers in the only
last 2 years and an order of magnitude increase in publication rate
compared to first decade of LGS.  The science impact of LGS AO has
been significant in a small but important subset of fields and is
gaining use in other areas.  The Keck LGS system has provided most of
these current science gains, and we can eagerly expect a similarly
large leap as the Gemini-North and VLT systems come into regular
science use.  From both qualitative and quantitative considerations,
it is clear that the era of regular and important science from LGS AO
has arrived.

\begin{table}[h]
  \vskip -0.1in
  \caption{{\bf Published Keck LGS AO Science Performance.}}
\label{table:pubished}
\begin{center}       
\vskip -1ex
\begin{tabular}{p{1.8in}|p{1.4in}|p{2.6in}|c}
         & Measurement  &                &     \\
Property & Uncertainty ($\lambda\lambda$)  & Science Object & Ref \\
\hline
\hline

\rule[-0.5ex]{0pt}{3.5ex}

Photometry: Relative  & $\approx$0.05~mag ($K$)  & Grav.\ lens, 2\arcsec\ size    & \citenum{2007ApJ...671.1196M} \\
                      & $\approx$0.03~mag ($K$)  & Wide (11\arcsec) binary        & \citenum{2007ApJ...664.1167K} \\
                      & $\approx$0.14~mag ($H$)  & High-$z$ supernova, $H\sim24$  & \citenum{2007AJ....133.2709M} \\
                      & $\approx$0.15~mags ($K$) & Grav. lens w/2 field stars     & \citenum{2007MNRAS.378..109M}  \\ 
                      & 0.06--0.11~mags ($JHK$)  & Galactic high energy source     & \citenum{2007ApJ...665L.135C}\\ 
                      &  &  & \\ 

\rule[-0.5ex]{0pt}{3.5ex}  
Photometry: Absolute  & n/a &  & \\ 
                      &  &  & \\ 

\hline
\rule[-0.5ex]{0pt}{3.5ex}  
Astrometry: Relative & $\approx$1.5~mas ($K$)    & Grav.\ lens, 2\arcsec\ size  & \citenum{2007MNRAS.378..109M}  \\
                     & $\sim$5~mas ($K$)         & Wide (11\arcsec) binary      & \citenum{2007ApJ...664.1167K}  \\ 
                     & 40--90~mas ($H$)          & Galactic high energy source   & \citenum{2007ApJ...665L.135C}\\ 
                     &  &  & \\ 

\rule[-0.5ex]{0pt}{3.5ex}  
Astrometry: Absolute & 0.01\arcsec\ ($K$) & Resolved galaxy + SN            & \citenum{2007ApJ...656..372G} \\
                     & 0.1\arcsec\ ($K$)  & Galactic high energy source     & \citenum{2007ApJ...665L.135C} \\
                     & 0.01\arcsec ($K$)  & Resolved galaxy + SN            & \citenum{2005ApJ...630L..29G} \\
                     &  &  & \\ 

\hline
\rule[-0.5ex]{0pt}{3.5ex}  
Binary (relative) photometry & $\approx$0.01~mag ($JHK$)     & Binary: $\{\approx$0.2\arcsec, $\Delta{mag}\approx0.3$\}  & \citenum{liu08-2m1534orbit}   \\
                             & $\approx$0.05~mag ($HK$)      & Binary: $\{\approx$0.1\arcsec, $\Delta{mag}\approx0.3$\}  & \citenum{liu08-2m1534orbit}   \\
                             & $\approx$0.3~mag ($K$)        & Binary: $\{\approx$0.5\arcsec, $\Delta{mag}\approx4.2$\}  & \citenum{2006ApJ...639L..43B} \\
                             & $\approx$0.1--0.3 mag ($K$)   & Binary: $\{\approx$1.2\arcsec, $\Delta{mag}\approx1.4$\}  & \citenum{2006ApJ...639L..43B} \\ 
                             & 0.04--0.09~mag ($JHK$)        & Binary: $\{\approx$1.2\arcsec, $\Delta{mag}\approx1.4$\}  & \citenum{2006ApJ...639L..43B} \\ 
                             & 0.05--0.06~mag ($JHK$)        & Binary: $\{\approx$0.07\arcsec, $\Delta{mag}\approx1.2$\} & \citenum{2006ApJ...649..389P} \\  
                             & 0.05~mag       ($JK$)         & Binary: $\{\approx$0.37\arcsec, $\Delta{mag}\approx1.5$\} & \citenum{2006ApJ...649..389P} \\ 
                             & 0.02--0.04~mag ($JHK$)        & Binary: \{0.29\arcsec, $\Delta{mag}\approx0.5$\}            & \citenum{2005astro.ph..8082L} \\ 
                             & 0.04--0.05~mag ($JHK$)        & Binary: \{0.11\arcsec, $\Delta{mag}\approx0.5$\}            & \citenum{2006astro.ph..5037L} \\ 
                             &  &  & \\ 

\rule[-0.5ex]{0pt}{3.5ex}  
Binary (relative) astrometry & $\approx$0.3--1.5~mas ($JHK$) & Binary: $\{\approx$0.2\arcsec, $\Delta{mag}\approx0.3$\}    & \citenum{liu08-2m1534orbit} \\
                             & $\approx$0.5--1.0~mas ($HK$)  & Binary: $\{\approx$0.1\arcsec, $\Delta{mag}\approx0.3$\}    & \citenum{liu08-2m1534orbit} \\
                             & $\approx$10--50~mas ($HK$)    & Binary: $\{\approx$0.5\arcsec, $\Delta{mag}\approx4.2$\}    & \citenum{2007Sci...316.1585B} \\
                             & $\approx$50 mas ($K$)         & Binary:  \{0.07\arcsec, $\Delta{mag}\approx0.9$\}           & \citenum{2007AJ....133.2320S} \\ 
                             & 4 mas ($JHK$)                 & Binary: $\{\approx$1.2\arcsec, $\Delta{mag}\approx1.4$\}    & \citenum{2006ApJ...639L..43B} \\ 
                             & 2.5 mas ($H$)                 & Binary: $\{\approx$0.07\arcsec, $\Delta{mag}\approx1.2$\}   & \citenum{2006ApJ...649..389P} \\  
                             & 3 mas ($K$)                   & Binary: $\{\approx$0.37\arcsec, $\Delta{mag}\approx1.5$\}   & \citenum{2006ApJ...649..389P} \\ 
                             & $\approx$20--30 mas ($K$)     & Binary: $\{\approx$0.6--1.3\arcsec, $\Delta{mag}\approx4$\} & \citenum{2005ApJ...632L..45B} \\ 
                             & 2 mas ($K$)                   & Binary: \{0.29\arcsec, $\Delta{mag}\approx0.5$\} & \citenum{2005astro.ph..8082L} \\ 
                             & 5 mas ($H$)                   & Binary: \{0.11\arcsec, $\Delta{mag}=0.7$\}            & \citenum{2006astro.ph..5037L} \\ 
                             &  &  & \\ 

\hline
\rule[-0.5ex]{0pt}{3.5ex}  
Crowded field astrometry     & 2 mas ($K$)                   & Star cluster: \{$K<20.5$~mag, $\Sigma_\star\approx 4/\Box\arcsec$\}      & \citenum{2008ApJ...675.1278S} \\ 
                             &  &  & \\ 

\rule[-0.5ex]{0pt}{3.5ex}  
Crowded field relative phot  & $<$0.01 mag ($HK$)            & Stellar pops: \{$HK\lesssim19$~mag, $\Sigma_\star\approx 4/\Box\arcsec$\}   & \citenum{2007ApJ...662..272V} \\
                             & $\approx$0.05~mag ($HK$)      & Star cluster: \{$HK\lesssim19$~mag, $\Sigma_\star\approx 4/\Box\arcsec$\}   & \citenum{2007ApJ...662..272V} \\ 
                             & $\approx$0.05~mag ($HK$)      & Gal Center: \{$HK\lesssim$19~mag, $\Sigma_\star\approx 4/\Box\arcsec$\} & \citenum{2005ApJ...635.1087G} \\ 
                             &  &  & \\ 

\rule[-0.5ex]{0pt}{3.5ex}  
Crowded field absolute phot  & $\le0.04$~mag ($H,K$)         & Stellar pops: \{$HK \lesssim 22$~mag, $\Sigma_\star\approx 4/\Box\arcsec$\} & \citenum{2007ApJ...662..272V} \\
                             & 0.05~mag ($H$)                & Stellar pops: \{$H \lesssim 20$~mag, $\Sigma_\star\approx 3/\Box\arcsec$\}  & \citenum{2006astro.ph..4551S} \\
\hline
\end{tabular}
\end{center}
\end{table}

\acknowledgments     

We thank Claire Max, James Graham, Don Gavel, Scot Olivier, Bruce
Macintosh and the LLNL AO group for stimulating our early interest in
doing science with LGS AO.  We also have been fortunate to benefit
from the exceptional dedication and expertise of current and past Keck
Observatory staff for their support of our brown dwarf science
observations presented herein.  We thank Peter Wizinowich for
providing a bibliography of Keck LGS science papers, and Jean-Rene
Roy, Joe Jensen, Antonin Bouchez, and Marcus Kaspar for news about the
Gemini-North, Palomar, and VLT LGS systems.  MCL's research presented
herein is partially supported from NSF grant AST-0507833 and an Alfred
P. Sloan Research Fellowship.  We wish to recognize and acknowledge
the very significant cultural role and reverence that the summit of
Mauna Kea has always had within the indigenous Hawaiian community.  We
are most fortunate to have the opportunity to conduct observations
from this mountain.

\clearpage


\end{document}